\documentclass[12pt]{article}
\usepackage{amssymb,amsmath,epsfig}
\begin{document}
\title{\bf Static Spherically Symmetric Wormholes in $f(R,T)$ Gravity}

\author{M. Zubair $^{a}$ \thanks{mzubairkk@gmail.com; drmzubair@ciitlahore.edu.pk},
Saira Waheed $^{b}$ \thanks{swaheed@pmu.edu.sa} and Yasir Ahmad
$^{a}$ \thanks{yasirahmad6667@gmail.com} \\\\
$^{a}$ Department of Mathematics, COMSATS \\Institute Of Information
Technology
Lahore-54590, Pakistan.\\
$^{b}$ Prince Mohammad Bin Fahd University,\\
Al Khobar, 31952 Kingdom of Saudi Arabia.}

\date{}
\maketitle
\begin{abstract}
In this work, we explore wormhole solutions in $f(R,T)$ theory of
gravity, where $R$ is the scalar curvature and $T$ is the trace of
stress-energy tensor of matter. To investigate this, we consider
static spherically symmetric geometry with matter contents as
anisotropic, isotropic and barotropic fluids in three separate
cases. By taking into account Starobinsky $f(R)$ model , we analyze
the behavior of energy conditions for these different kind of
fluids. It is shown that the wormhole solutions can be constructed
without exotic matter in few regions of spacetime. We also give the
graphical illustration of obtained results and discuss the
equilibrium picture for anisotropic case only. It is concluded that
the wormhole solutions with anisotropic matter are realistic and
stable in this gravity.
\end{abstract}

{\bf PACS:}\\

\section{Introduction:}

After Edwin Hubble's theory of expanding universe, current
observations from Supernovae Type Ia and CMBR (Cosmic Microwave
Background Radiations) \cite{1}, have confirmed the phenomena of
accelerated expanding universe. The modified theories are quite
useful in present era because these theories can help to explain the
possible cosmic expansion history and its related concepts. In this
context, $f(R)$ theory is appeared as one of the first and simplest
modifications to the Einstein-Hilbert action. This theory has been
extensively employed to discuss the dark energy (DE) and mainly the
accelerating cosmic expansion \cite{2}. Furthermore, $f(R)$ theory
of gravitation provides us the scenarios of early time inflation and
late time expansion of the accelerated universe \cite{3}. The
discussions about DE and late time cosmic acceleration are also
explained in some other modified theories of gravity such as
$f(\tau)$ (where ``$\tau$" being the torsion) \cite{4}, Gauss-Bonnet
Gravity \cite{7}, Brans-Dicke theory \cite{8} and $f(T,T_G)$,
\cite{9} etc.

Few years ago, Harko et al. \cite{10} introduced a modification to
Einstein's gravity and named it as $f(R,T)$ theory of gravity. This
was basically an extension to $f(R)$ gravity obtained by introducing
the trace ``$T$" of the energy-momentum tensor together with the
Ricci scalar ``$R$". Furthermore, they derived corresponding field
equations from the coupling of matter and geometry in metric
formalism for some specific cases. Recently, Houndjo \cite{11}
reconstructed some cosmological models of the form
$f(R,T)=f_1(R)+f_{2}(T)$, in the presence of auxiliary scalar field
with two known examples of scale factor that correspond to an
expanding universe. In \cite{12}, authors considered cosmological
scenarios based on $f(R,T)$ theories of gravity and numerically
reconstructed the function $f(R,T)$ for holographic DE model that
can reproduce the same expansion history as generated in general
relativity (GR). Till present time, different cosmological aspects
have been addressed in $f(R,T)$ gravity including reconstructions
schemes, anisotropic solutions, energy conditions, thermodynamics,
viscous solutions, phase space perturbations and stability, etc.
\cite{13}.

Wormholes are hypothetical topological features that provide a
subway for different space times apart from each other. In 1935,
Einstein and Rosen \cite{15} firstly obtained the wormhole solutions
known as Lorentzian wormholes or Schwarzchild wormholes. On the
basis of nature, wormholes are of two kinds: static wormholes and
dynamic wormholes. Normally, an exotic fluid is needed for the
formation of static wormholes which violates the NEC in GR. Lobo and
Oliveira \cite{16} explained the fact that how wormhole solutions
can be formed without violation of energy conditions, $i.e., $ WEC
and NEC, in $f(R)$ theory of gravity. They reconstructed $f(R)$ by
considering trace-less fluid and equations of state for some
particular shape function, to discuss the evolution of energy
conditions.

In \cite{17}, the behavior of ordinary matter was studied to check
whether it can support wormholes in $f(R)$ theory. For this purpose,
WEC and NEC were analyzed in anisotropic, barotropic and isotropic
fluids and it was observed that the barotropic fluid satisfies these
conditions in some certain regions of the space-time while for the
other two fluids, these conditions were violated. So, wormhole
solutions can be obtained without exotic matter in few regions of
space-time only, without violating the energy conditions which are
necessary for the existence of wormhole solutions in GR \cite{18}.
Recently, the wormhole geometries are studied in $f(R,T)$ gravity
\cite{19} by taking a particular equation of state (EoS) for the
matter field into account. They showed that effective stress-energy
is responsible for violation of the NEC.

Here, we are interested to find wormhole solutions by introducing
additional matter contributions in $f(R)$ model(without involving
any form of exotic matter). We analyze the behavior of shape
function, WEC and NEC to explore the suitable regions for existence
of wormhole solutions using anisotropic, barotropic and isotropic
fluids. This paper has following sequence. In section \textbf{2}, we
present a short introduction of $f(R,T)$ gravity by developing the
field equations. Section \textbf{3} relates to the discussion of
wormhole geometries in $f(R,T)$  theory of gravity for three types
of fluids. Finally, section \textbf{4} comprises of concluding
remarks.

\section{$f(R,T)$ Gravity}

Here, we will give a short introduction to $f(R,T)$ theory of
gravity. In his pioneer work, Harko et al. presented a new
generalization of $f(R)$ gravity by taking a coupling of Ricci
scalar with matter field into account as follows \cite{10}
\begin{equation}\label{1*}
\mathcal{I}=\int{dx^4\sqrt{-g}\left[f(R,T)+\mathcal{L}_{\mathrm{m}}\right]}.
\end{equation}
In above action, $f(R,T)$ represents a generic function of Ricci
scalar $R$ and the energy-momentum tensor trace $T=T^\mu_\mu$. Here,
in the action, we have assumed the gravitational units, i.e., $c =
8\pi{G} =1$ and also matter ingredients are introduced by the
Lagrangian density $\mathcal{L}_{(matter)}$. This theory is
considered as more successful as compared to $f(R)$ gravity in the
sense that such theory can include quantum effects or imperfect
fluids that are neglected in simple $f(R)$ generalization of GR. The
metric $g_{\mu\nu}$ variation of the above action leads to the
following set of field equations:
\begin{eqnarray}\label{1}
R_{{\mu}{\nu}}f_{R}(R,T)&-&\frac{1}{2}g_{{\mu}{\nu}}f(R,T)+\big(g_{{\mu}{\nu}}
{\Box}-{\nabla}_{\mu}{\nabla}_{\nu}\big) f_{R}(R,T)\nonumber\\
&=&T_{{\mu}{\nu}} -f_{T}(R,T) \Theta_{{\mu}{\nu}}-f_{T}(R,T)
T_{{\mu}{\nu}}.
\end{eqnarray}
This set involves derivative operators like $\nabla$ and $\Box$ that
represent covariant derivative and four-dimensional Levi-Civita
covariant derivative also known as d'Alembert operator,
respectively. Also, the notations $f_{R}(R,T)$ and $f_{T}(R,T)$
correspond to derivatives of $f(R,T)$ with respect to Ricci scalar,
i.e., $\frac{\partial{f(R,T)}}{\partial{R}}$ and energy-momentum,
i.e., trace $\frac{\partial{f(R,T)}}{\partial{T}}$, respectively.
The term $\Theta_{{\mu}{\nu}}$ is defined by
\begin{equation}\nonumber
\Theta_{{\mu}{\nu}}=\frac{g^{\alpha{\beta}}{\delta}T_{{\alpha}{\beta}}}
{{\delta}g^{\mu{\nu}}}=-2T_{{\mu}{\nu}}+g_{\mu\nu}\mathcal{L}_m
-2g^{\alpha\beta}\frac{\partial^2\mathcal{L}_m}{{\partial}
g^{\mu\nu}{\partial}g^{\alpha\beta}},
\end{equation}
where the matter energy-momentum tensor is introduced which is given
by the following equation \cite{28}
\begin{equation}\label{2*}
T_{{\mu}{\nu}}^{(m)}=-\frac{2}{\sqrt{-g}}\frac{\delta(\sqrt{-g}
{\mathcal{\mathcal{L}}_{m}})}{\delta{g^{{\mu}{\nu}}}}=g_{{\mu}{\nu}}
\mathcal{L}_{m}-\frac{2{\partial}{\mathcal{L}_{m}}}
{\partial{g^{{\mu}{\nu}}}}.
\end{equation}
Here the second part of the above equation can be obtained, if the
matter Lagrangian is assumed to depend only on the metric tensor
rather than on its derivatives.

The source of anisotropic fluid is defined by the following energy
momentum tensor
\begin{equation*}
T_{{\mu}{\nu}}=({\rho}+p_r)V_{\mu}V_{\nu}-p_{t}g_{{\mu}{\nu}}+(p_r-p_t)\chi_{\mu}\chi_{\nu},
\end{equation*}
where $V_{\mu}$ is the 4-velocity of the fluid defined as
$V^{\mu}=e^{-a}\delta^{\mu}_{0}$  satisfying $V^{\mu}V_{\mu}=1$ and
$\chi^{\mu}=e^{-b}\delta^{\mu}_{1}$ gives $\chi^{\mu}\chi_{\mu}=-1$.
Herein, we choose $\mathcal{L}_{(matter)}={\rho}$, then the
expression for $\Theta_{{\mu}{\nu}}$ takes the following form
\begin{equation}\nonumber
\Theta_{{\mu}{\nu}}=-2T_{{\mu}{\nu}}+{\rho}g_{{\mu}{\nu}}.
\end{equation}

Consequently, the field equations (\ref{1}) can be expressed as effective
Einstein field equations of the form
\begin{equation}\label{2}
R_{{\mu}{\nu}}-\frac{1}{2}R g_{{\mu}{\nu}} =T_{{\mu}{\nu}}^{eff},
\end{equation}
where $T_{{\mu}{\nu}}^{eff}$ is the effective energy-momentum tensor
in $f(R,T)$ gravity which is defined by
\begin{eqnarray}\nonumber
{T}_{{\mu}{\nu}}^{eff}&=&\frac{1}{f_{R}(R,T)}\left[(1+f_T(R,T))T_{\mu\nu}
- \rho g_{\mu\nu}f_T(R,T)+\frac{1}{2}(f(R,T)\right.\\
&-&\left.Rf_{R}(R,T))g_{\mu\nu}+
({\nabla}_{\mu}{\nabla}_{\nu}-g_{{\mu}{\nu}}{\Box})f_{R}(R,T)\right].\label{3}
\end{eqnarray}

\section{Wormhole Geometries with Three Different Matter Contents}

In this section, we will discuss static spherically symmetric
wormholes with three types of matter contents: anisotropic,
isotropic and barotropic. Consider a line element that describes
static spherically symmetric geometry of the form
\begin{equation}\label{7}
ds^2=e^{a(r)}dt^2-e^{b(r)}dr^2-r^2(d\theta^2+sin^2\theta d\phi^2),
\end{equation}
where $a(r)$ is an arbitrary function of $r$ and for the wormhole
geometry, we have $e^{-b(r)}=1-\beta(r)/r$. The terms $a(r)$ and
$\beta(r)$ represent the redshift function, and the shape function,
respectively \cite{18}. For the surface vertical to the wormhole
throat, we must have a minimum radius at $r = \beta(r_0) = r_0$,
then it increases from $r_0$ to $r\rightarrow\infty$. An important
condition to have a typical wormhole solution is the flaring out
condition of the throat, given by $\frac{(\beta-\beta'r)}{\beta^2} >
0$ and moreover, $\beta(r)$ needs to meet the condition
$\beta'(r_0)<1$ that is imposed at the throat $\beta(r_0) = r =
r_0$. In GR, these conditions hints the existence of exotic form of
matter which requires the violation of the NEC. Also, the condition
$1-\beta(r)/r>0$ needs to be satisfied.

The field equations can be rearranged to find the expressions for $\rho$,
$p_r$ and $p_t$ as follows
\begin{eqnarray}\nonumber
\nonumber\rho&=&\frac{1}{e^{b}}\left[\bigg(\frac{a'}{r}-\frac{a'b'}{4}+\frac{a''}{2}+\frac{a'^2}{4}\bigg)f_{R}(R,T)+
\bigg(\frac{b'}{2}-\frac{2}{r}\bigg)f'_{R}(R,T)-f''_{R}(R,T)\right.\\\label{4}
&-&\left.\frac{f(R,T)}{2}e^{b}\right], \end{eqnarray}
\begin{eqnarray}\nonumber
p_r &=&
\frac{1}{e^{b}(1+f_{T}(R,T))}\left[\bigg(\frac{b'}{r}-\frac{a'b'}{4}-\frac{a''}{2}-\frac{a'^2}{4}\bigg)
f_{R}(R,T)+\bigg(\frac{a'}{2}+\frac{2}{r}\bigg)f'_{R}(R,T)\right.\\
&&\left.+\frac{f(R,T)}{2}e^{b}\right]\label{5}
-\frac{{\rho}f_{T}(R,T)}{(1+f_{T}(R,T))},\\\nonumber
\nonumber p_t &=& \frac{1}{e^{b}(1+f_{T}(R,T))}\left[\bigg(\frac{(b'-a')r}{2}-e^{b}+1\bigg)\frac{f_{R}(R,T)}{r^2}+
\bigg(\frac{a'-b'}{2}+\frac{1}{r}\bigg)f'_{R}(R,T)\right.\\
&&\left.+f''_{R}(R,T)+\frac{f(R,T)}{2}e^{b(r)}\right]
-\frac{{\rho}f_{T}(R,T)}{(1+f_{T}(R,T))}.\label{6}
\end{eqnarray}
It can be observed that the above equations appeared as much
complicated to find the explicit expressions of $\rho$, $p_r$ and
$p_t$, since $f(R,T)$ has direct dependence on trace of
stress-energy tensor. In this scenario, we find that the only
possibility left is to choose the function as $f(R,T)=f(R)+f(T)$
with $f(T)=\lambda T$, where $\lambda$ being the coupling parameter.
Here, we set this choice for $f(R,T)$ and simplify the above
equations (\ref{4})-(\ref{6}) as follows
\begin{eqnarray}\label{8**}
\rho&=&\frac{1}{2(1+2{\lambda})}\bigg[\frac{2+5{\lambda}}{(1+{\lambda})}Z_{1}+{\lambda}Z_{2}
+2{\lambda}Z_{3}\bigg],\\\label{9**}
p_r&=& \frac{-1}{2(1+2{\lambda})}\bigg[\frac{{\lambda}}{(1+{\lambda})}Z_{1}-(2+3{\lambda})Z_{2}
+2{\lambda}Z_{3}\bigg],\\\label{10*}
p_t&=&\frac{-1}{2(1+2{\lambda})}\bigg[\frac{{\lambda}}{(1+{\lambda})}Z_{1}-(2+3{\lambda})Z_{2}
+2{\lambda}Z_{3}\bigg],
\end{eqnarray}
where
\begin{eqnarray}\nonumber
Z_1&=&\frac{1}{e^{b}}\bigg[\bigg(\frac{a'}{r}-\frac{a'b'}{4}+\frac{a''}{2}+\frac{a'^2}{4}\bigg)f_{R}+
\bigg(\frac{b'}{2}-\frac{2}{r}\bigg)f'_{R}-f''_{R}-\frac{f}{2}e^{b}\bigg],
\\\nonumber
Z_2&=&\frac{1}{e^{b}(1+{\lambda})}\bigg[\bigg(\frac{b'}{r}+\frac{a'b'}{4}-\frac{a''}{2}-\frac{a'^2}{4}\bigg)
f_{R}+\bigg(\frac{a'}{2}+\frac{2}{r}\bigg)f'_{R}+\frac{f}{2}e^{b}\bigg],
\\\nonumber
Z_3&=&\frac{1}{e^{b}(1+{\lambda})}\bigg[\bigg(\frac{(a'-b')r}{2}-e^{b}+1\bigg)\frac{-f_{R}}{r^2}+
\bigg(\frac{a'-b'}{2}+\frac{1}{r}\bigg)f'_{R}\\\nonumber
&&+f''_{R}+\frac{f}{2}e^{b}\bigg].
\end{eqnarray}
%

In \cite{19*}, authors have presented the study of energy conditions
in $f(R,T)$ gravity. We recommend the readers to see these papers
for having an overview of this subject. As for the other modified
theories, the violation of the NEC in $f(R,T)$ gravity imposes the
condition $T_{\mu\nu}^{eff}\kappa^{\mu}\kappa^{\nu} < 0$, that is,
$\rho^{eff}+p^{eff}<0$. Here, we find the following expression:
\begin{equation}\nonumber
\rho^{eff}+p_r^{eff}=\frac{1}{f_R}(\rho+p_r)(1+\lambda)+\frac{1}{f_R}\left(1-\frac{\beta}{r}\right)
\left(f''_{RR}+f'_{R}\frac{\beta-\beta'r}{2r^2(1-\frac{\beta}{r})}\right).
\end{equation}
Using the field equations, this leads to
\begin{equation}\nonumber
\rho^{eff}+p_r^{eff}=\frac{1}{r^3}\bigg({\beta}'r-\beta\bigg),
\end{equation}
which is similar to that in $f(R)$ gravity. Here, if we use the
flaring out condition $\frac{(\beta-\beta'r)}{\beta^2} > 0$, it
results in $\rho^{eff}+p^{eff}<0$. In this case, we have kept NEC
satisfied for matter energy tensor, however the additional curvature
components which arise due to modification of Einstein's gravity
play role for violation of the NEC.

In this study, we take a specific $f(R)$ model representing $R^n$ extension
of well known Starobinsky model \cite{3} and is given by \cite{19**}
\begin{equation}\nonumber
f(R)=R+\alpha R^2+\gamma R^n,
\end{equation}
where $n\geqslant3$, $\alpha$ and $\gamma$ are arbitrary constants.
The choice of $\alpha=\gamma=0$ implies the $\Lambda$ correction to
GR. Basically, we want to take power law model that should be
singularity free as well as it should be the generalization of
linear models that are used in most of the literature for wormhole
discussions. In literature \cite{2s}, it is pointed out that the
power law models are always of great interest, e.g., $f(R)=\xi R^n$
with $\xi,n$ are any constants. In this model, there exists big rip
singularities for negative range of $n$. They also argued that if we
impose $n>1$ with positive $\xi$, then $f(R)\rightarrow\infty$ only
when $R\rightarrow\infty$. Thus under these conditions, the possible
presence of singularity can be avoided. Further, in literature,
another form of Starobinsky model with disappearing cosmological
constant is defined $f(R)=R+\lambda
R_0[(1+\frac{R^2}{R_0})^{-n}-1]$. Clearly, this model suffers the
singularity problem. However, they also claimed that this
singularity can be cured by adding a term $\propto R^2$ \cite{3s}.
It can be easily seen that our used Starobinsky model has different
form involving one $R^2$ term, therefore our used model does not
suffer with any singularity problem. One can explore another viable $f(R)$ model named as Hu-Sawicki model to 
present interesting cosmic features \cite{4s}.

For this model, the field equations (\ref{8**})-(\ref{10*}) take the
form
\begin{eqnarray}\nonumber
\rho&=&\frac{e^{-b}}{4r^2(1+{\lambda})(1+2{\lambda})}\bigg[(1+2{\alpha}R+
n{\gamma}R^{n-1})(2r^2{a''}(1+2{\lambda})-
{a'}r(1+2{\lambda})\\\nonumber
&\times&\big(r{b'}-4)+r^2{a'}^2(1+{\lambda})+4r{b'}{\lambda}+4{\lambda}(e^{b}-1))+
(2{\alpha}+n(n-1){\gamma}R^{n-2})\\\nonumber
&\times&\big(3r^2{a'}{\lambda}+r(2+3{\lambda})(r{b'}-4)R'-2r^2{R}e^{b}(1+{\lambda})
(R+{\alpha}R^2+{\gamma}R^n)-2r^2\\\label{11}
&\times&\big(2+3{\lambda})({\gamma}n(n-1)(n-2)R^{n-3}R'^2+2{\alpha}R''+{\gamma}n(n-1)R^{n-2}R''\bigg],\\\nonumber
p_r &=&\frac{e^{-b}}{4 r^2 (1+\lambda ) (1+2 \lambda )}
\bigg[\big(1+2 \alpha  R+n \gamma  R^{n-1}\big) \big(-4
\big(-1+e^{b}\big) \lambda +r\big(\\\nonumber &-&(r+2 r \lambda )
a'^2+4 (1+\lambda ) b'+r (1+2 \lambda ) a' b'-2 r (1+2 \lambda )
a''\big)\big)+r \big(\big(2\alpha\\\nonumber &+&n(n-1) \gamma
R^{n-2}\big) \big(8+12 \lambda +r (2+\lambda ) a'+r \lambda b'\big)
R'+2 r \big(e^{b} (1+\lambda ) \big(R\\\label{12} &+&\alpha
R^2+\gamma  R^n\big)-2 \alpha  \lambda  R''-n(n-1) \gamma  \lambda
R^{n-3} \big((n-2) R'^2+R R''\big)\big)\big)\bigg],\\\nonumber p_t
&=&\frac{e^{-b}}{4 r^2 \big(1+3 \lambda +2 \lambda ^2\big)} \bigg[2
\big(1+2 \alpha  R+n \gamma  R^{n-1}\big) \big(2 \big(e^{b}-1\big)
(1+\lambda )\\\nonumber &-&(r+2 r \lambda ) a'+r b'\big)+r
\big(\big(2 \alpha +n(n-1) \gamma  R^{n-2}\big) \big(4 (1+\lambda
)+r (2+\lambda ) a'\\\nonumber &-&r (2+3 \lambda ) b'\big) R'+2 r
\big(e^{b} (1+\lambda ) \big(R+\alpha  R^2+\gamma  R^n\big)+(2+3
\lambda ) \big(n(n-1)\\\label{13} &\times&(n-2)\gamma  R^{n-3}
R'^2+2 \alpha  R''+n(n-1)\gamma  R^{n-2}
R''\big)\big)\big)\bigg].\\\nonumber
\end{eqnarray}
In further discussion, we take a particular value for $n$ as $n=3$.
Moreover, the redshift function is chosen to be constant with
$a'(r)=0$. In coming sections, we discuss the energy bounds for
three different fluid configurations.

\subsection{Anisotropic fluid}

Initially, we consider the anisotropic fluid model with the following choice
of shape function \cite{20}-\cite{23}
\begin{equation}\label{1a}
b(r)=-\text{ln}\bigg[1-\bigg(\frac{r_0}{r}\bigg)^{m+1}\bigg],
\end{equation}
where $m$ and $r_0$ are arbitrary constants. Since
$e^{-b(r)}=1-\beta(r)/r$, so, in our case, Eq.(\ref{1a}) implies the
following form of shape function
\begin{equation}\label{zz}
\beta(r)=\frac{(r_0)^{m+1}}{r^m}.
\end{equation}
Clearly, $\beta(r)$ is characterized on the basis of $m$ and can
result in different forms, which have been explored in literature as
shown in Table \ref{Table1}. Here, $\beta(r)$ satisfies the
necessary conditions for the existence of shape function. To meet
the flaring out condition $\beta'(r)<1$, one need to set $m<1$.
Also, the constraint $\beta(r_0)=r_0$ is trivially satisfied.
Moreover, this shape function also satisfies the condition for
asymptotically flat spacetime, i.e.,
$\beta(r)/r=r_0^{1-m}r^{m-1}\rightarrow0$ as $r\rightarrow\infty$.

\begin{table}[!ht]
    \begin{tabular}{|c|c|c|c|c|c|c|}
        \hline
        $m$ & $m=1$ & $m=1/2$ & $m=1/5$ & $m=0$& $m=-1/2$ & $m=-3$\\
        \hline \hline

{Shape Functions $\beta(r)$}  & $r_0^2/r$ & $r_0\sqrt{r_0/r}$ & $r_0^{6/5}r^{-1/5}$ & $r_0$ & $\sqrt{r_0r}$ & $r_0^2r^3$   \\ \cline{2-4}
       \hline
       \end{tabular}
    \caption{Shape functions corresponding to different choices of parameter $m$.}
    \label{Table1}
\end{table}

In \cite{17}, Lobo and Oliveira discussed the wormhole geometries in
$f(R)$ gravity using the above defined shape function for the
choices: $m=1$ and $m=-1/2$. Recently, Pavlovic and Sossich
\cite{22} discussed the existence of wormholes without exotic matter
in different $f(R)$ models employing this shape function (\ref{1a})
with $m=1/2$.

Substituting $b(r)$ in Eqs. (\ref{11})-(\ref{13}), we obtain
\begin{eqnarray}
\nonumber\rho&=&\frac{r_0 m}{r^9
(1+\lambda)(1+2\lambda)}\left(\frac{r_0}{r}\right)^m \bigg[-r^4
\bigg(r^2 (1+2 \lambda )+2 (2+m) (3+m) \alpha  (2+3
\lambda)\bigg)\\\nonumber &+& r_0 \left(\frac{r_0}{r}\right)^m r
\bigg(-12 m (3+m) (5+2 m) \gamma  (2+3 \lambda ) +r^2 \alpha
\bigg(30+26 m+6 m^2+45\lambda\\\nonumber &+&9m \lambda (4+m)
\bigg)\bigg)+2 \gamma r_0^2 m \big(\frac{r_0}{r}\big)^{2 m}
\bigg(m( 154+226 \lambda) +(15 m^2+99)\\
&\times& (2+3 \lambda )\bigg)\bigg],\\\label{7a}\nonumber
p_r&=&\frac{r_0}{r^9 (1+\lambda ) (1+2 \lambda )}
\left(\frac{r_0}{r}\right)^m \bigg[-r^4 \bigg(r^2 (1+2 \lambda )+2
m\alpha (3+m)(4+10\lambda+m \lambda)\bigg)
\end{eqnarray}
\begin{eqnarray}\nonumber &+&2 r_0^2 m^2 \left(\frac{r_0}{r}\right)^{2 m}
\gamma \bigg(33 (2+7 \lambda )+m (20+(122+15 m) \lambda )\bigg)+r_0
m \left(\frac{r_0}{r}\right)^m r \bigg(-12 m\gamma\\\label{7b}
&\times&(3+m)  (4+(13+2 m) \lambda )+r^2 \alpha  \bigg(20+55 \lambda
+m (6+(28+3 m) \lambda )\bigg)\bigg)\bigg],\\\nonumber
p_t&=&\frac{r_0 }{2 r^9 (1+\lambda ) (1+2 \lambda
)}\left(\frac{r_0}{r}\right)^m \bigg[-4 r_0^2 m^2
\left(\frac{r_0}{r}\right)^{2 m} \gamma  \bigg(231+169 m+30
m^2+(3+m)
\\\nonumber &\times&(121+45 m) \lambda \bigg)+r^4 \bigg((1+m) r^2
(1+2 \lambda)+4 m (3+m) \alpha(6+10 \lambda +m (2\\\nonumber
 &+&3 \lambda ))\bigg)+2 r_0 m \left(\frac{r_0}{r}\right)^m r \bigg(-r^2 \alpha  \bigg(40+65 \lambda +m (16+3
 m) (2+3 \lambda )\bigg)\\
 &+&12 m\gamma (3+ m)\bigg(12+19 \lambda +m (4+6 \lambda )\bigg)\bigg)\bigg].\label{7c}
\end{eqnarray}

In the following discussion, we present the suitable choice of
parameters for the viability of WEC: $\rho>0$ and NECs:
$\rho+p_r>0$, $\rho+p_t>0$. We compare the different shape functions depending on the choice
of parameter $m$.

\begin{itemize}
  \item $\beta(r)=r_0\sqrt{\frac{r_0}{r}}$
\end{itemize}
Here, we fix $r_0=1$ and $m=1/2$ and
discuss the viability ranges of $\alpha$, $\gamma$ and $r$ for two
cases of coupling constant, $\lambda>-1$ and $\lambda<-1$. In case of
WEC, we find the following constraints:
\begin{itemize}
\item For $\lambda<-1$, WEC is valid if $r\geq3$ and $\alpha\geq15$,
here $r$ depends on the choice of $\alpha$, for very large $\alpha$,
we can increase the validity region. However, $r$ obeys the initial
bound $r\geq1.3$ for greater values of $\alpha$. In left plot of the
Fig. \textbf{1}, we show the evolution of WEC versus $\alpha$,
$\gamma$ and $r$ for $\lambda=-2$. One can see that there are some
small regions where WEC is also valid for $\alpha<0$. For small
choice of $r$, we refer the readers to see the right plot in Fig.
\textbf{1}. We have shown the plot for $\lambda=-2$, it can be seen
that there some small regions of validity involving $\alpha<0$ and
very small range of $r$.
\begin{itemize}
\item For small region like $0<r\leq1$, we require $\gamma\leq-30$
and for $1<r<3$, we require $\alpha>0$ with $\gamma\geq30$.
\end{itemize}
\item For $\lambda>-1$, the validity of WEC needs $\alpha\leq-20$ and
$r\geq2.8$. In left plot of Fig. \textbf{2}, we have shown the
validity regions for $\lambda=2$, it can be seen that there some
small regions of validity involving $\alpha>0$ and very small range
of $r$. In right plot of Fig. \textbf{2}, we show the evolution for
small ranges of $r$ and find the following constraints:
\begin{itemize}
\item For small region like $0<r<1$, we require $\gamma\geq20$ and
for $1.2\leq r<2.8$, the validity needs negative values of both
$\alpha$ and $\gamma$.
\end{itemize}
\end{itemize}
\begin{figure}\centering
\epsfig{file=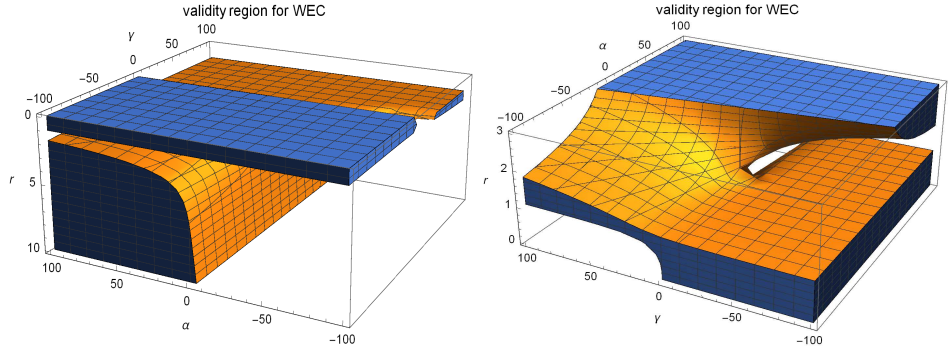, width=0.45\linewidth, height=1.8in}
\epsfig{file=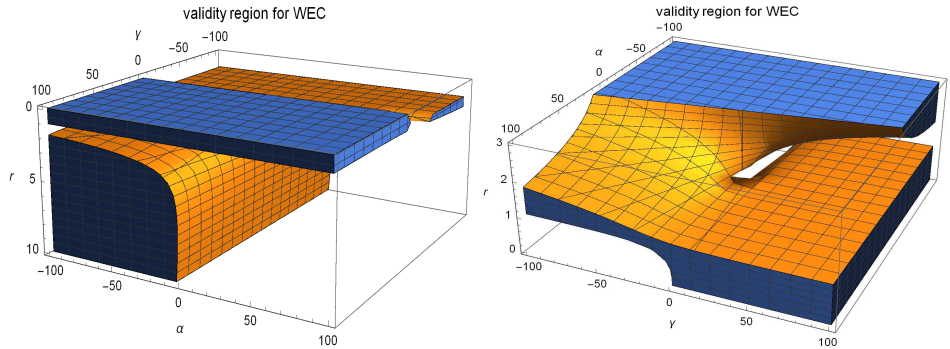, width=0.45\linewidth, height=1.8in}
\caption{Validity of WEC for $\lambda=-2$ with $c=1$ and $m=1/2$. In
right plot, we present the evolution for small $r$ which is as clear
in left plot. }
\end{figure}
\begin{figure}\centering
\epsfig{file=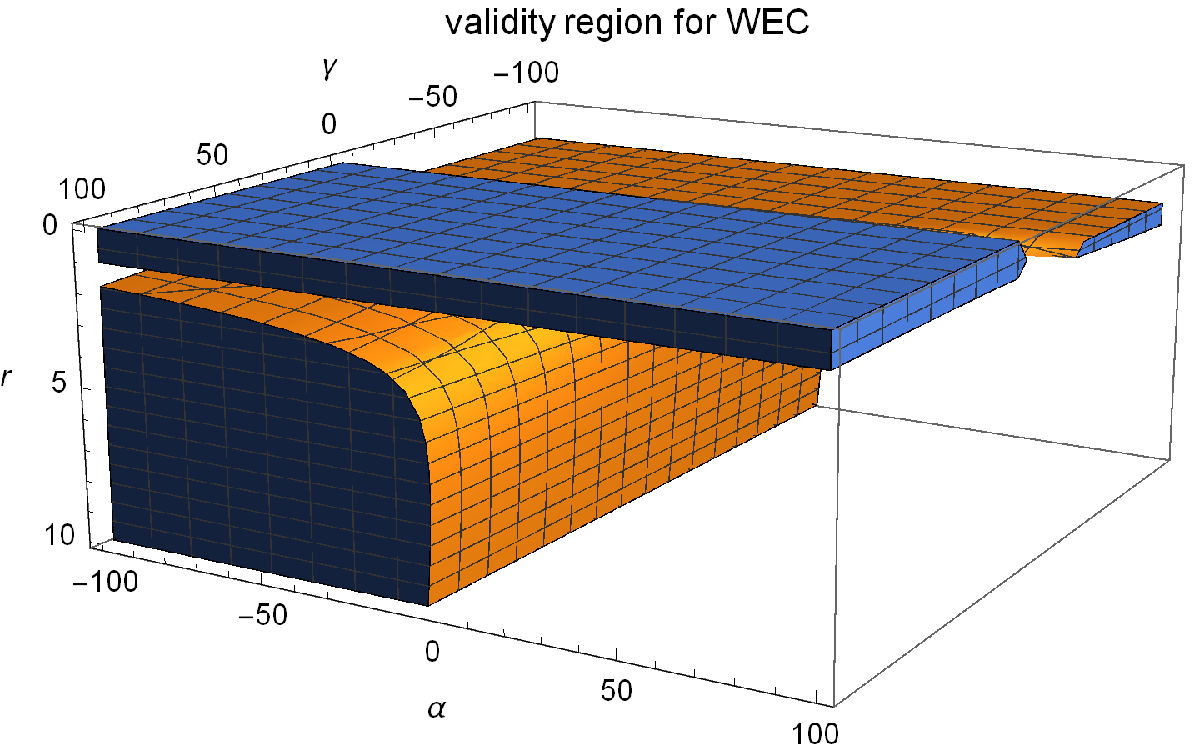, width=0.45\linewidth, height=1.8in}
\epsfig{file=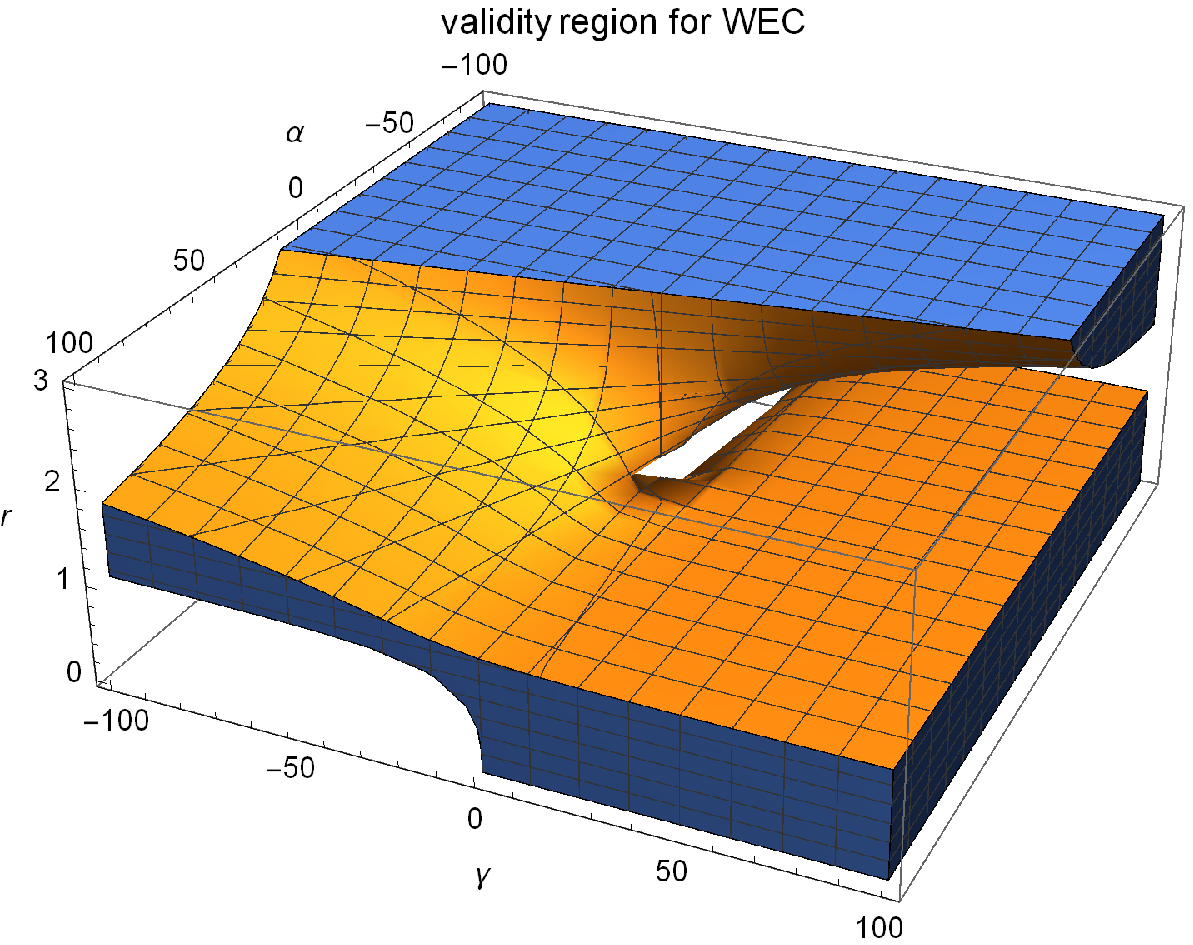, width=0.45\linewidth, height=1.8in}
\caption{Validity of WEC for $\lambda=-2$ with $c=1$ and $m=1/2$. In
right plot, we present the evolution for small $r$ which is as clear
in left plot.}
\end{figure}
Now we discuss the validity regions for $\rho+p_r>0$ and
$\rho+p_t>0$. Again we develop two cases depending on the choice of
$\lambda$.
\begin{itemize}
\item For $\lambda<-1$, $\rho+p_r>0$ is valid in following regions:\\
$0 < r < 1$ with $\gamma\leq-13$; $r\geq1.1$ with $\alpha>0$ and
$\gamma>-1$; $r\geq3$ with $\alpha\geq10$.
\item The validity of $\rho+p_t>0$ can be met for three cases, i.e.,\\
$0<r<1$ with $\gamma\geq15$; $r\geq1.1$ with $\alpha\leq-10$, and
$\gamma<0$; if $r\geq3$ with $\alpha\leq-10$.
\end{itemize}
Now we present the constraints for $\lambda>-1$.
\begin{itemize}
\item  Here, $\rho+p_r>0$ can be satisfied for four different ranges
depending on the choice of $r$: if $0<r<1$ with $\gamma\geq15$; if
$r\geq 2.9$, $\alpha\leq-15$; if $1<r<2.9$, $\alpha<0$, $\gamma<0$
and if $r\geq1.1$ with $\gamma<0$ and $\alpha\leq-10$.
\item In case of $\rho+p_t>0$, we can find the validity for the
following choices of the parameters: For $0 < r < 1$ with $\gamma
\leq-16$; for $r \geq2.9$ with $\alpha\geq10$; for $1 \leq r < 2.9$
with $\alpha>0$, $\gamma > 0$ and for $r \geq 1.1$ with $(\alpha,
\gamma)> 0$.
\end{itemize}
In Figs. \textbf{3} and \textbf{4}, we present the evolution of
$\rho+p_r>0$ and $\rho+p_t>0$ for $\lambda=2$ and $\lambda=-2$,
respectively. We find that there is no region of similarity between
$\rho+p_r>0$ and $\rho+p_t>0$, though one can find the same validity
range for both $\rho>0$ and $\rho+p_r>0$. In Fig. \textbf{5}, we
show the plots of the $\rho$, $p_r$ and $p_t$ for $c=1$, $n=0.5$,
$\lambda=2$, $\alpha=-2$ and $\gamma=-0.1$. It can be seen that both
$\rho>0$ and $p_r>0$ are satisfied but $p_t>0$ is violated. Thus, in
anisotropic case, the normal matter threading the wormhole does not
satisfy $\rho+p_t>0$.
\begin{figure}\centering
\epsfig{file=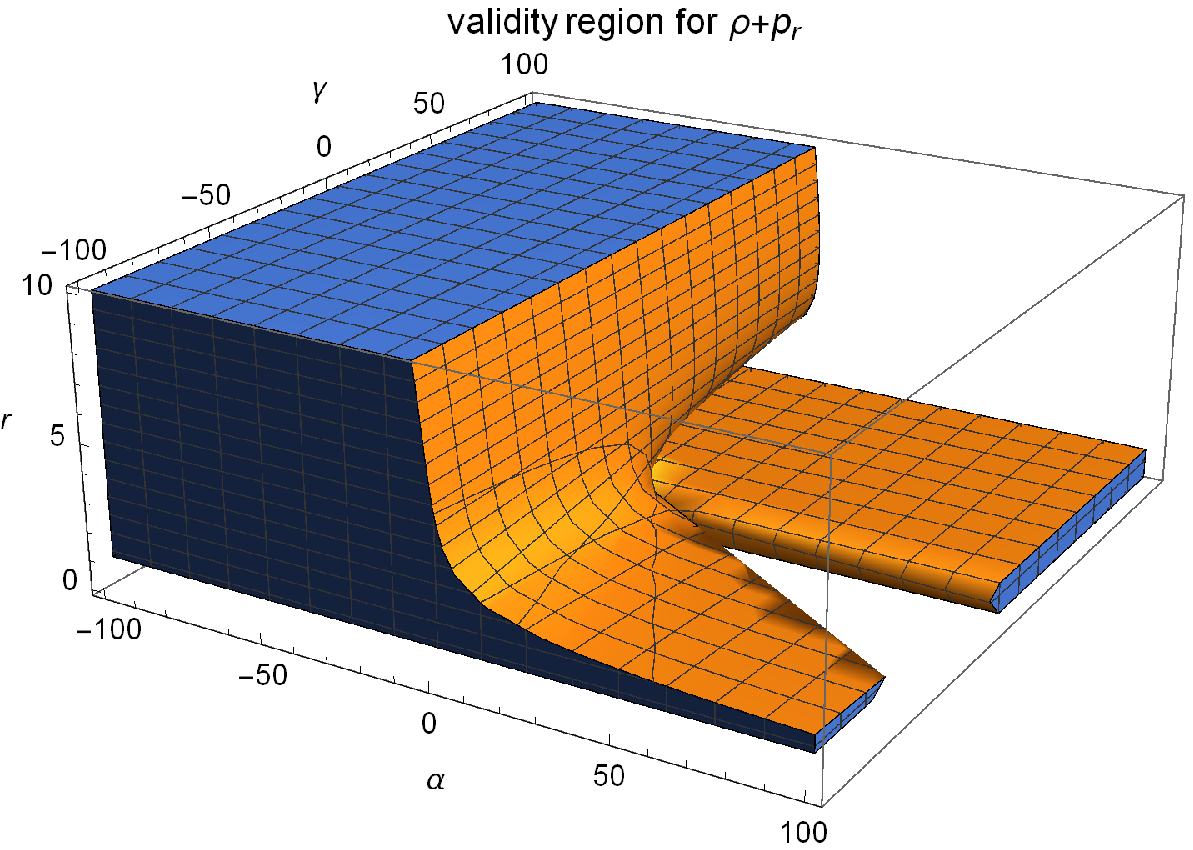, width=0.45\linewidth, height=1.8in}
\epsfig{file=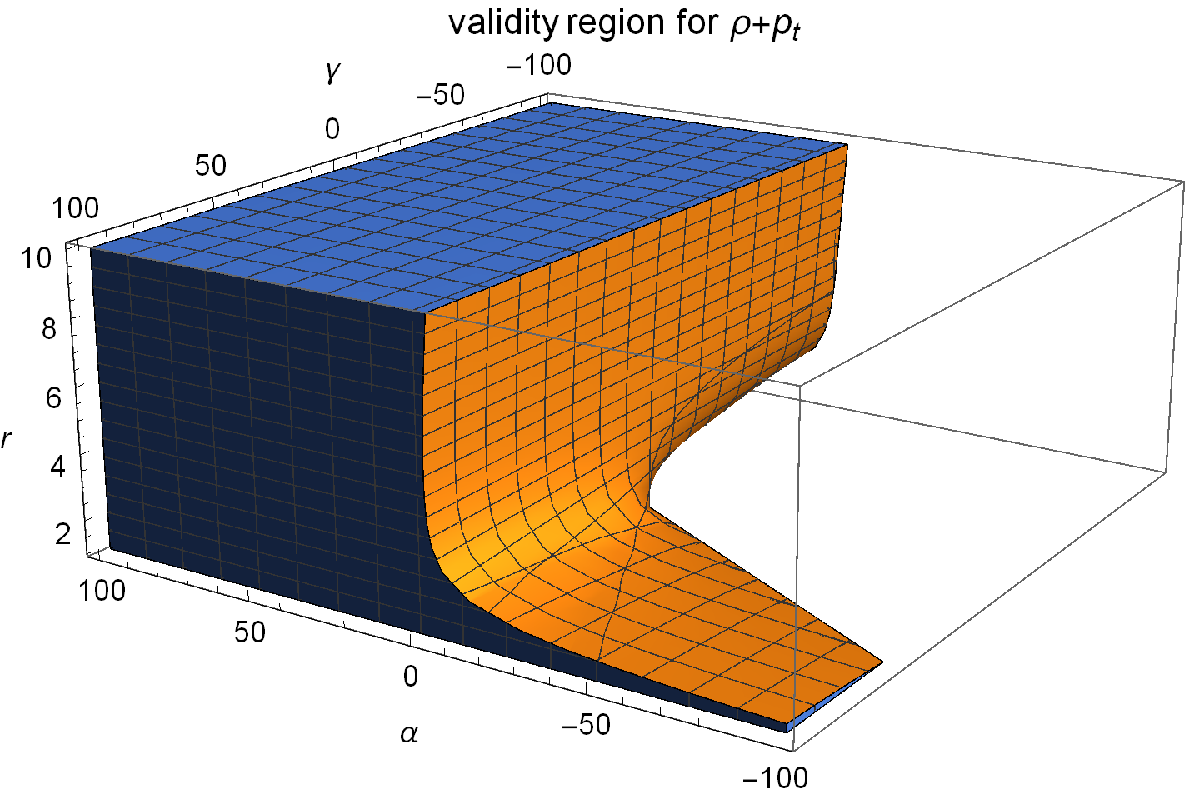, width=0.45\linewidth, height=1.8in}
\caption{Validity of $\rho+p_r>0$ and $\rho+p_t>0$ for $\lambda=2$
with $c=1$ and $m=1/2$.}
\end{figure}
\begin{figure}\centering
\epsfig{file=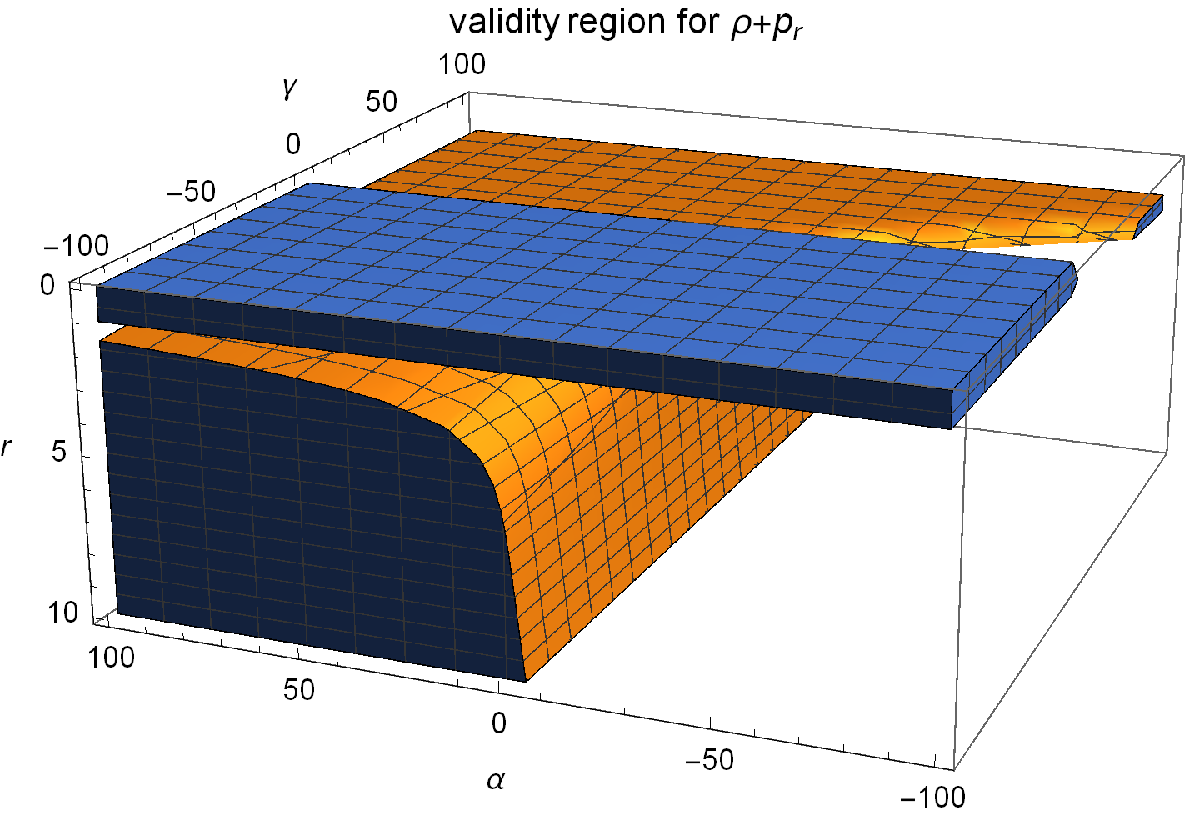, width=0.45\linewidth, height=1.8in}
\epsfig{file=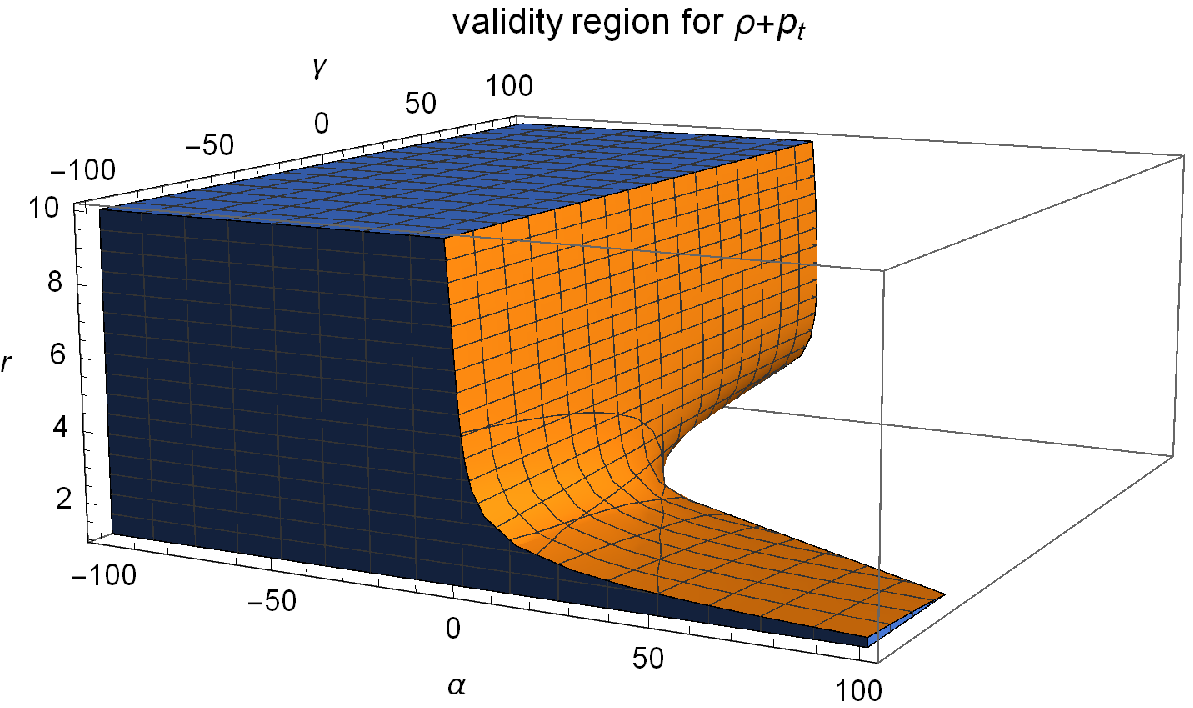, width=0.45\linewidth, height=1.8in}
\caption{Validity of $\rho+p_r>0$ and $\rho+p_t>0$ for $\lambda=-2$
with $c=1$ and $m=1/2$.}
\end{figure}
\begin{figure}\centering
\epsfig{file=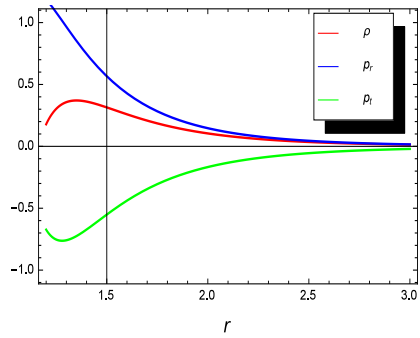, width=0.45\linewidth, height=1.8in}
\caption{Evolution of $\rho$, $p_r$ and $p_t$ for the anisotropic
case with $\lambda=2$.}
\end{figure}
\begin{itemize}
\item $\beta(r)=r_0^2/r$
\end{itemize}
For this choice of shape function, one has to set $m=1$. The results for this choice are
very similar as the inequalities remain the same, with only difference with the bounds of parameters.
\begin{itemize}
\item {For $\lambda<-1$, WEC is valid if $r\geq3$ and $\alpha\geq20$.
For small region like $0<r\leq1$, we require $\gamma\leq-10$
and for $1<r<3$, we require $\alpha>0$ with $\gamma\geq10$.
$\rho+p_r>0$ is valid in following regions:\\
$0 < r < 1$ with $\gamma\leq-15$; $r\geq1.1$ with $\alpha>0$ and
$\gamma>0$; $r\geq3$ with $\alpha\geq15$.\\
The validity of $\rho+p_t>0$ can be met for three cases, i.e.,\\
$0<r<1$ with $\gamma\geq15$; $r\geq1.1$ with $\alpha\leq-5$, and
$\gamma<0$; if $r\geq3$ with $\alpha\leq-15$.

\item For $\lambda>-1$, the validity of WEC needs $\alpha\leq-20$ and
$r\geq3$. For small region like $0<r<1$, we require $\gamma\geq10$ and
for $1.1\leq r<3$, the validity needs negative values of both
$\alpha$ and $\gamma$, i.e, $\alpha\leq-5$ and $\gamma\leq-20$.
NEC with radial pressure can be satisfied for the following ranges
depending on the choice of $r$: if $0<r<1$ with $\gamma\geq15$; if
$r\geq 3$, $\alpha\leq-15$; if $1<r<3$, $\alpha<-5$, $\gamma<0$.

In case of $\rho+p_t>0$, we can find the validity for the
following choices of the parameters: For $0 < r < 1$ with $\gamma
\leq-16$; for $r \geq3$ with $\alpha\geq10$; for $1 \leq r < 3$
with $\alpha>0$, $\gamma > 0$.}
\end{itemize}
We would like to mention here that all the choices like $m=1, 1/2,
-1/2, 1/5$ \cite{20}-\cite{23}, implies the same sort of results as
presented in detail for the case of $m=1/2$. However, the parameter
$m=-3$ gives significantly different results.

\begin{itemize}
  \item $\beta(r)=r_0^2r^3$
\end{itemize}
For $m=-3$, we find the shape function of the form
$\beta(r)=r_0^2r^3$, in this case $\rho$, $p_r$ and $p_t$ appears to
be independent of $r$. Here, the choice of $\lambda<-1$ results in
following constraints: WEC, $\rho+p_r>0$ and $\rho+p_t>0$ are valid
only if $\gamma\leq-20$ for all values of $\alpha$. If one sets
$\lambda>-1$ then the energy conditions are valid only if
$\gamma\geq15$ for all values of $\alpha$. Hence, for this choice
the normal matter threading the wormhole is on cards.

\subsection{Equilibrium Condition}

Now we present some discussion about the equilibrium picture of
wormhole solutions. For wormhole solutions, the equilibrium picture
can be discussed by taking Tolman-Oppenheimer-Volkov equation given
by:
\begin{equation}\label{8a}
\frac{dp_r}{dr}+\frac{\sigma'}{2}(\rho+p_r)+\frac{2}{r}(p_r-p_t)=0,
\end{equation}
that is defined for the metric:
\begin{eqnarray}\nonumber
ds^2=e^{\sigma(r)}dt^2-e^bdr^2-r^2(d\theta^2+\sin^2\theta d\phi^2),
\end{eqnarray}
where $\sigma(r)=2a(r)$. This equation describes the equilibrium
picture by considering anisotropic force arising from anisotropic
matter, hydrostatic and gravitational forces that are identified as
follows
\begin{eqnarray}\nonumber
F_{gf}=-\frac{\sigma'(\sigma+p_r)}{2},\quad
F_{hf}=-\frac{dp_r}{dr},\quad F_{af}=2\frac{(p_t-p_r)}{r},
\end{eqnarray}
and the equilibrium equation takes the form:
\begin{eqnarray}\nonumber
F_{hf}+F_{gf}+F_{af}=0.
\end{eqnarray}
In our case, since we have taken $a'(r)=0$, therefore $F_{gf}$ turns
out to be zero and hence the previous equation takes the form:
$$F_{hf}+F_{af}=0.$$
In our case, these forces takes the following form:
\begin{eqnarray}\nonumber
F_{hf}&=&\frac{7}{2}r^{-9/2}+(\frac{11}{2})(\frac{59.5}{3})\alpha
r^{-13/2}-\frac{21}{12}(515.5\gamma)r^{-23/2}+\frac{3}{2}(504\gamma)r^{-10}\\\label{8b}
&-&\frac{7}{6}(116.5\alpha)r^{-8},\\\nonumber
F_{af}&=&\frac{2}{r}[\frac{681.5\gamma}{6}r^{-21/2}-\frac{4.5}{6}r^{-7/2}-\frac{112\alpha}{6}r^{-11/2}+\frac{125\alpha}{6}r^{-7}
-\frac{630\gamma}{6}r^{-9}-r^{-7/2}\\\label{8b}
&-&\frac{59.5\alpha}{3}r^{-11/2}+\frac{515.5\gamma}{6}r^{-21/2}-\frac{1}{6}r^{-9}(504\gamma-116.5\alpha
r^2)],
\end{eqnarray}
where we have used Eqs.(\ref{7a}), (\ref{7b}) and (\ref{7c}) with
$m=1/2,~n=3,~c=1$ and $\lambda=-2$. For the second case, i.e,
$\lambda=2$, these forces are given by:
\begin{eqnarray}\nonumber
F_{hf}&=&-\frac{35}{42}r^{-9/2}-\frac{11}{42}(87.5\alpha)r^{-13/2}+\frac{21}{84}(667.5\gamma)r^{-23/2}-\frac{672}{2}(9\gamma)r^{-10}\\\label{8c}
&+&\frac{7}{2}(162.5)r^{-8},\\\nonumber
F_{af}&=&\frac{2}{r}[-\frac{1}{30}(1327.5\gamma)r^{-21/2}+\frac{7.5}{30}r^{-7/2}+\frac{210\alpha}{30}r^{-11/2}-\frac{240\alpha}{30}r^{-7}
+\frac{1218\gamma}{30}r^{-9}\\\label{8d}
&+&\frac{5}{21}r^{-7/2}+\frac{87.5\alpha}{21}r^{-11/2}-\frac{667.5\gamma}{42}r^{-21/2}+\frac{672\gamma}{42}r^{-9}-\frac{162.5\alpha}{42}r^{-7}].
\end{eqnarray}
The graphical behavior of these forces is given in Fig. \textbf{6}.
Here we have taken $\alpha=15$ and $\gamma=30$ for which WEC is
compatible as discussed previously. The right plot corresponds to
the case $\lambda=2$, while left plot corresponds to $\lambda=-2$.
It can be seen that both these forces show similar behavior but
their behaviors are in opposite direction. Therefore, these forces
can cancel each other's effect and hence leads to the stability of
total configuration. Thus we can conclude that in case of $f(R,T)$
gravity with anisotropic matter, the wormhole solutions remain
stable.
\begin{figure}\centering
\epsfig{file=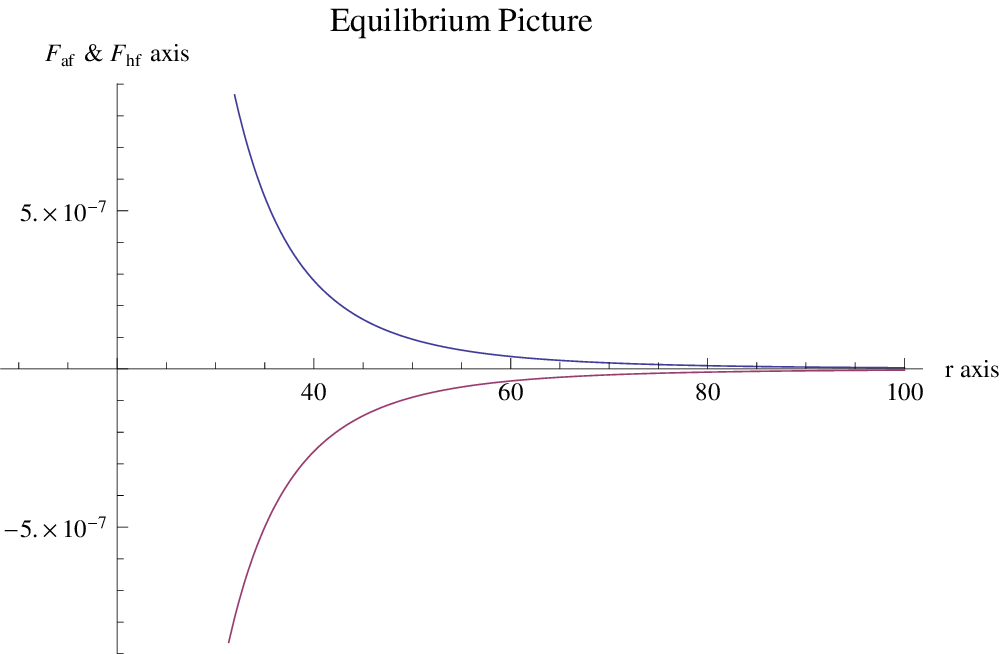, width=0.45\linewidth, height=2in}
\epsfig{file=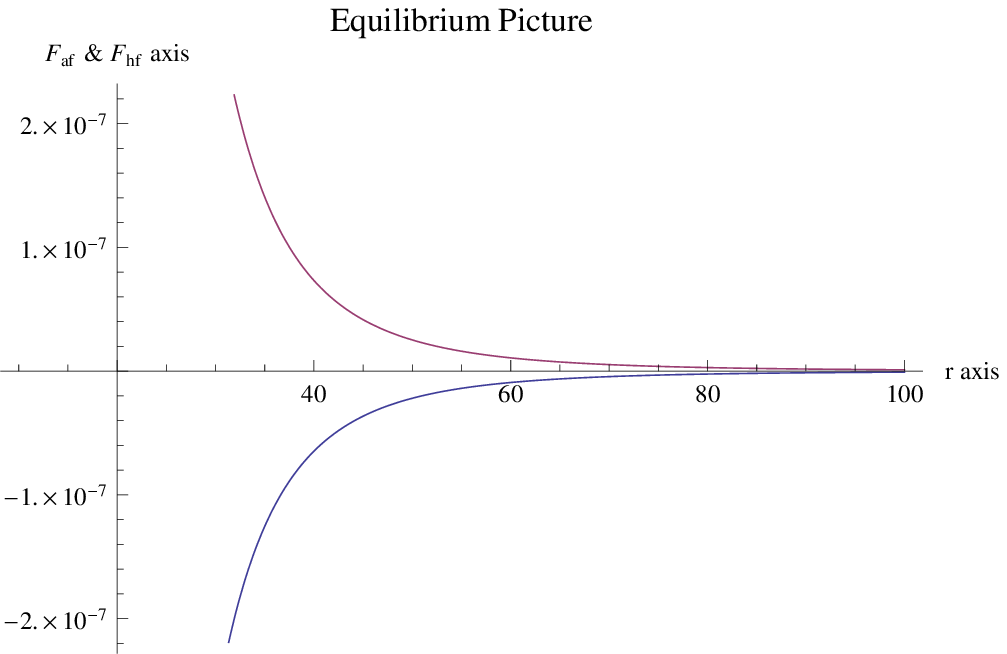, width=0.45\linewidth, height=2in}
\caption{Evolution of $F_{af}$ and $F_{hf}$ versus $r$. Herein, we
choose $m=1/2,~c=1,~n=3,~\alpha=15$ and $\gamma=30$. The right and
left plots correspond to $\lambda = 2$ and $\lambda = -2$,
respectively.}
\end{figure}

\subsection{Isotropic fluid}

For this case, we consider $p_r = p_t = p$. Hence, the isotropic condition
results in following equation
\begin{eqnarray}
\nonumber&&\frac{1}{r (1+\lambda )}e^{-b} \bigg[-12 r^3 \bigg(e^{b} \big(r^2 \alpha -6
\gamma \big)-\gamma \bigg) b'^3+120 r^4 \gamma  b'^4-12 r^2 b'^2 \bigg(e^{b}\\\nonumber
&\times&\bigg(r^2 \alpha-28 \gamma \bigg)+36 \gamma +22 r^2 \gamma  b''\bigg)+r b' \bigg(-e^{2 b} r^4+28
e^{b} r^2 \alpha -12 e^{2 b} r^2 \alpha \\\nonumber
&-&204 \gamma +120 e^{b} \gamma+84 e^{2 b} \gamma
+4 r^2 \bigg(7 e^{b} \big(r^2 \alpha -6 \gamma \big)+18 \gamma \bigg) b''+48 r^3
\gamma  b^{(3)}\bigg)\\\nonumber
&+&2 \bigg(\big(e^{b}-1\big) \bigg(-4 e^{b} \big(7 r^2 \alpha -66 \gamma \big)
-276 \gamma +e^{2 b} \big(r^4-4 r^2\alpha +12 \gamma \big)\bigg)\\\nonumber
&+&8 r^2 \bigg(e^{b} \big(r^2\alpha- 18\gamma \big)+18 \gamma \bigg) b''+24 r^4 \gamma
 b''^2-4 r^3 \bigg(e^{b} \big(r^2 \alpha -6\gamma \big)\\
&+&6 \gamma \bigg) b^{(3)}\bigg)\bigg]=0.\label{7*}
\end{eqnarray}
Here, one can present the above equation in terms of shape function
$\beta(r)$. It can be seen that Eq.(\ref{7*}) is highly non-linear,
which can not solved analytically. We use the numerical scheme to
solve the above equation and present the results in Figs.
\textbf{7}, \textbf{8} and \textbf{9}. In the left plot of Fig.
\textbf{7}, the evolution of shape function is shown which indicates
the increasing behavior and the condition $\beta(r)<r$ is obeyed,
whereas the right plot represents one of the fundamental wormhole
condition, i.e., the spacetime is asymptotically flat,
$\beta(r)/r\rightarrow0$ as $r\rightarrow\infty$. The throat is
located at $r_0=0.0932726$ so that $\beta(r_0)=r_0$. The derivative
of shape function is shown in right plot of Fig. \textbf{8}, it can
be seen that $\beta'(r_0)=0.0027559$ so that the condition
$\beta'(r_0)<1$ is satisfied. In right plot of Fig. \textbf{8}, we
plot the function $\beta(r)-r$, it is found that $\beta(r)-r<0$,
which validates the condition $1-\beta(r)/r>0$. The behavior of WEC
and NEC is shown in Figure \textbf{9}. It can be seen that $\rho>0$
throughout the evolution but $\rho+p>0$ can bet met in some
particular regions. Thus, a micro wormholes can be formed for this
case.
\begin{figure}\centering
\epsfig{file=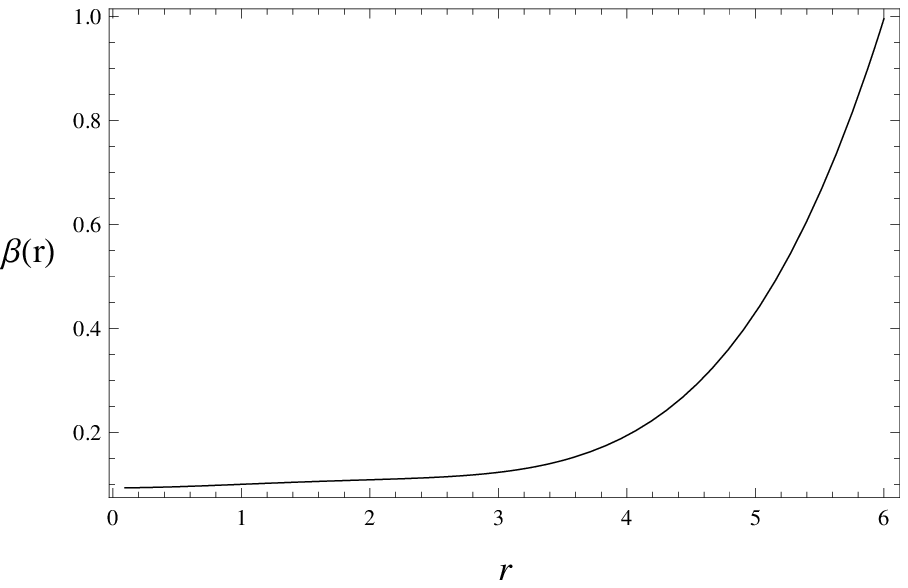, width=0.45\linewidth, height=2in}
\epsfig{file=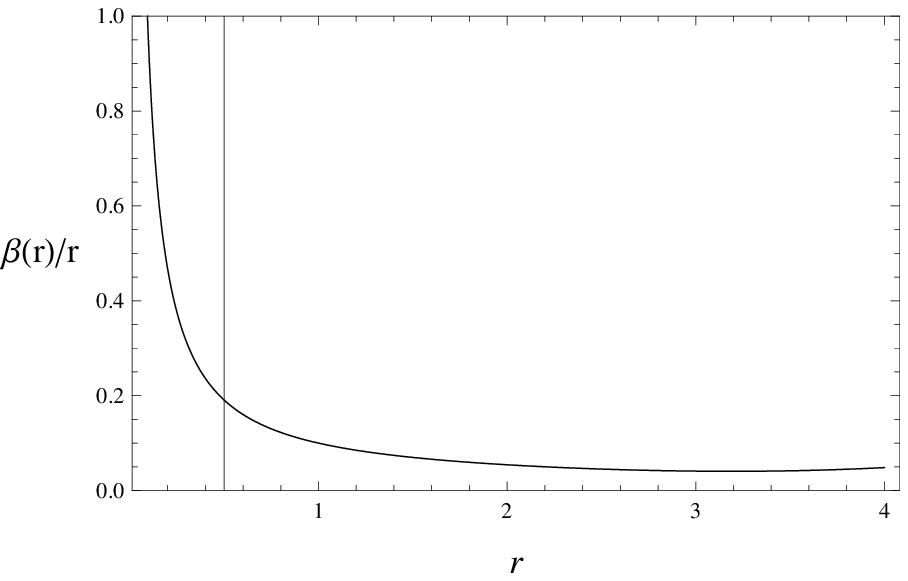, width=0.45\linewidth, height=2in}
\caption{Evolution of $\beta(r)$ and $\beta(r)/r$ versus $r$.
Herein, for isotropic case, we set $\lambda = 2, \alpha = 0.6,
\gamma=-0.2.$}
\end{figure}
\begin{figure}\centering
\epsfig{file=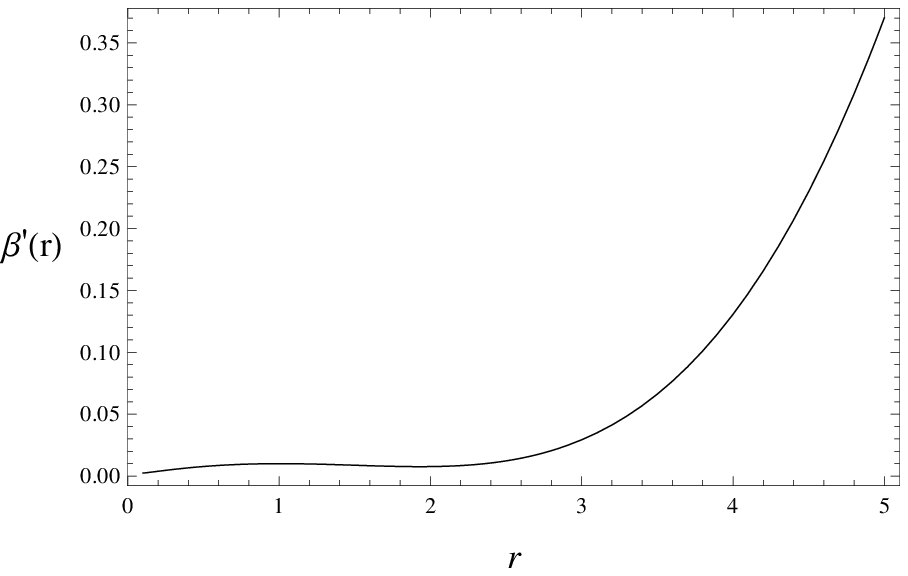, width=0.45\linewidth, height=2in}
\epsfig{file=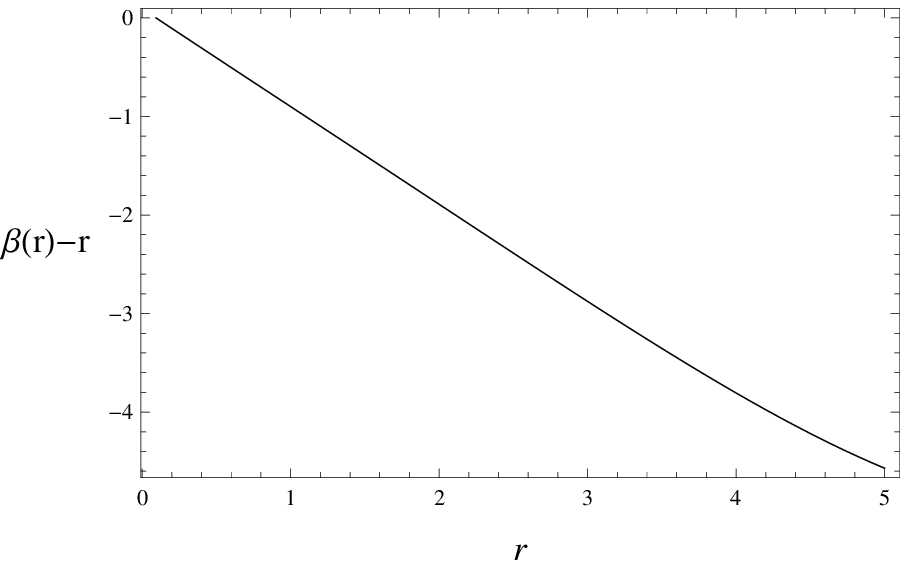, width=0.45\linewidth, height=2in}
\caption{Evolution of $\beta'(r)$ and $\beta(r)-r$ versus $r$.}
\end{figure}
\begin{figure}\centering
\epsfig{file=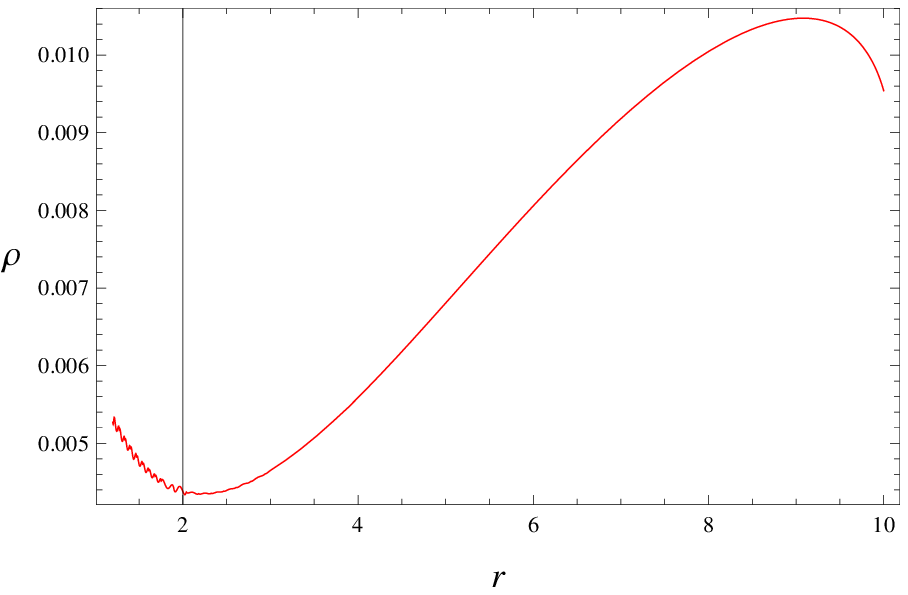, width=0.45\linewidth, height=2in}
\epsfig{file=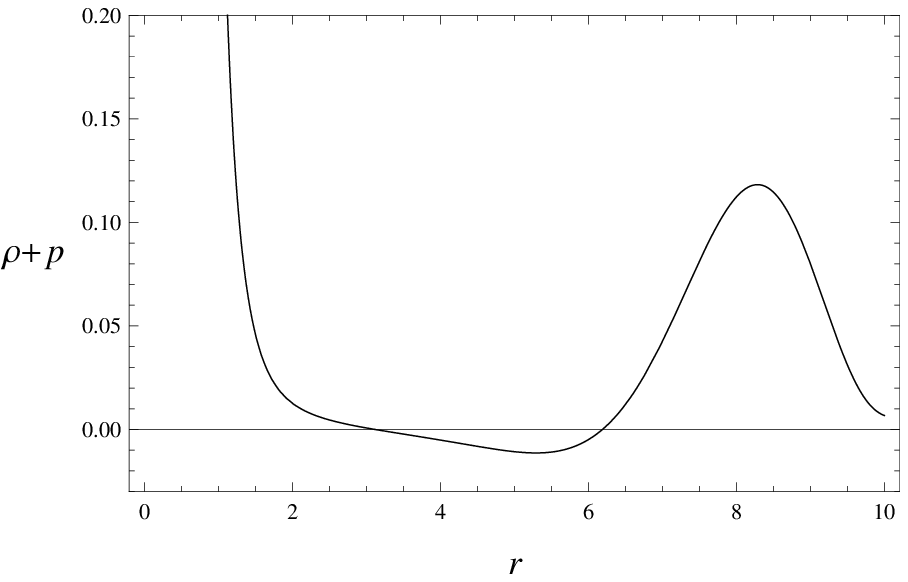, width=0.45\linewidth, height=2in}
\caption{Evolution of $\rho$ and $\rho+p$ for isotropic case.}
\end{figure}

\subsection{Specific EoS $p_r=k\rho$}

\begin{figure}\centering
\epsfig{file=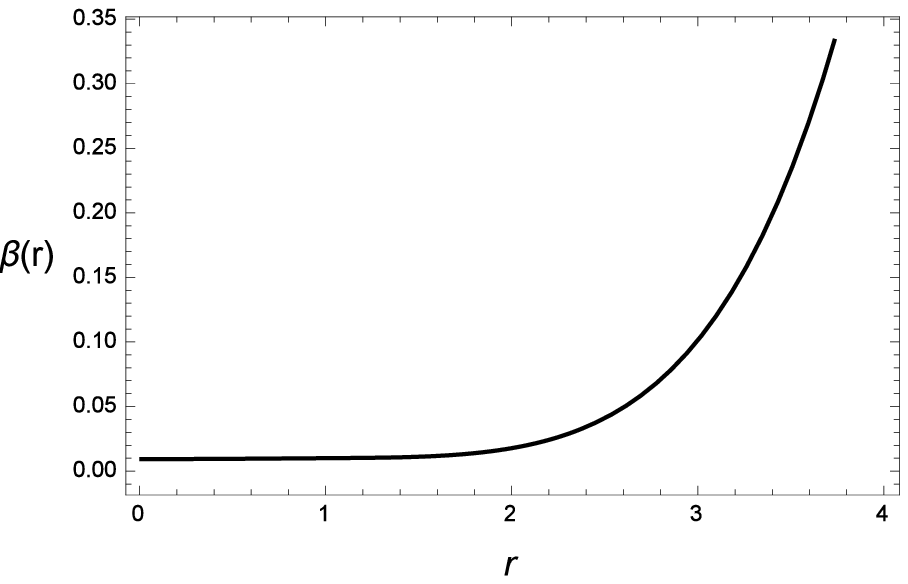, width=0.45\linewidth, height=2in}
\epsfig{file=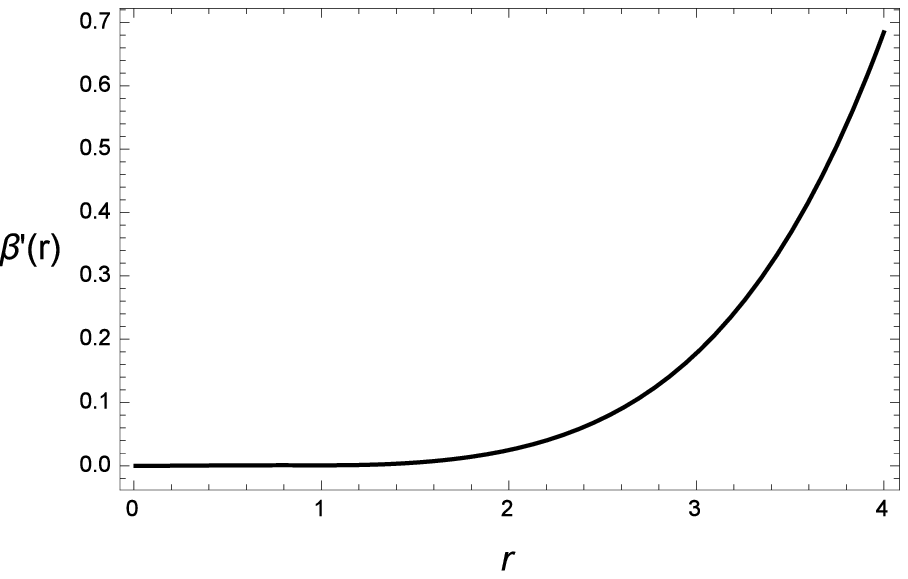, width=0.45\linewidth, height=2in}
\caption{Evolution of $\beta(r)$ and $\beta'(r)$ versus $r$. Herein,
for the
 EoS $p_r=k\rho$ we set $\lambda=-2, \alpha=0.1, \beta=-30, k=0.001$.}
\end{figure}
\begin{figure}\centering
\epsfig{file=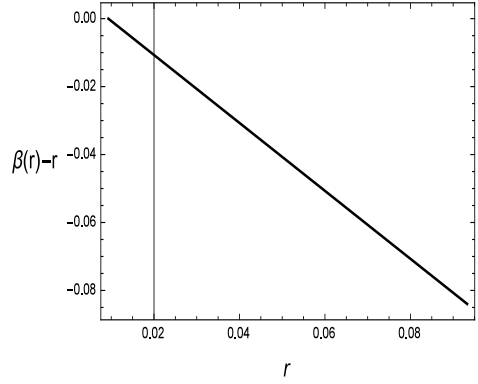, width=0.45\linewidth, height=2in}
\caption{Evolution of $\beta(r)-r$ for EoS $p_r=k\rho$.}
\end{figure}
In this case, we apply an EoS involving energy density and radial
pressure, i.e., $p_r=k\rho$. Such EoS has been applied in $f(R)$ and
$f(T)$ gravities \cite{17,20,21} to discuss the wormhole solutions.
Using the above defined EoS along with dynamical equation, we find
the following constraint to calculate the shape function.
\begin{eqnarray}\nonumber
&&\frac{1}{r (1+\lambda ) (1+2 \lambda )}e^{-b} \bigg[4 \gamma
(23+62 \lambda -k  (37+58 \lambda ))+e^{3 b} (1+k ) \bigg(r^4 (1+2
\lambda )\\\nonumber &-&2 r^2 (\alpha +3 \alpha  \lambda )+4 (\gamma
+4 \gamma  \lambda )\bigg)-2 e^{b} \bigg(-30 \gamma  (-3+5 k +8 (k
-1) \lambda )+r^2 \\\nonumber &\times& \alpha \big(-7-15 \lambda +k
(5+9 \lambda )\big)\bigg)-e^{2 b} \bigg(-12 r^2 \alpha(k-1 ) (1+2
\lambda )+r^4 (1+k )
\\\nonumber &\times&(1+2 \lambda )+12 \gamma  (k  (13+22
\lambda )-7-18 \lambda ) \bigg)+r \bigg(r^2\bigg(3 e^{b} \big(r^2
\alpha -6 \gamma \big)\bigg (k  (3 \lambda\\\nonumber &+&2)-\lambda
\bigg)+2 \gamma  \bigg(20+29 \lambda +k \big(32+53 \lambda
)\bigg)\bigg) b'^3+30 r^3 \gamma \bigg(\lambda -k  (2+3 \lambda
)\bigg) b'^4\\\nonumber &+&6 r b'^2 \bigg(2 \gamma(11 k-3 +14 (k-1 )
\lambda )-e^{b} \bigg(r^2 \alpha (1+k )(1+2 \lambda )+2 \gamma (7 k
-12 \\\nonumber &\times&\lambda+8 k \lambda-3 )\bigg) +11 r^2 \gamma
\bigg(k  (2+3 \lambda )-\lambda\bigg) b''\bigg)+b' \bigg(12
\big(e^{b}-1\big) \gamma \bigg(8+ k\big(e^{b}
\end{eqnarray}
\begin{eqnarray}\nonumber &-&1\big)(\lambda-1)+13 \lambda +3 e^{b} \lambda
\bigg)+e^{b} r^2 \bigg(-2 \alpha  \bigg( k (4+3 \lambda )-5
\lambda\bigg)+e^{b} \bigg(-6 \alpha (1\\\nonumber&+& k ) \lambda
+r^2(k +2 k \lambda )\bigg)\bigg)+r^2 \bigg(-7 e^{b} \big(r^2 \alpha
-6 \gamma \big) \bigg(k  (2+3 \lambda )-\lambda\bigg)-6 \gamma
\big(8\\\nonumber &+&7 \lambda +9 k(2+3 \lambda )\big)\bigg) b''+12
r^3 \gamma (\lambda -k (2+3 \lambda )) b^{(3)}\bigg)+2 r
\bigg(\bigg(6 \gamma (4+11 \lambda \\\nonumber &-&3 k  (2+3
\lambda))+e^{b} \bigg(r^2 \alpha  (4+2 k +7 \lambda +3 k \lambda )+6
\gamma (6 k-4 -11 \lambda +9 k  \lambda )\bigg)\bigg)\\\nonumber
&\times& b''+6 r^2 \gamma \bigg(\lambda-k  (2+3 \lambda )\bigg)
b''^2+r \bigg(e^{b} \big(r^2 \alpha -6 \gamma
\big)+6 \gamma \bigg) \bigg(-\lambda +k  (2 \\
&+& 3 \lambda\bigg ) b^{(3)}\bigg)\bigg)\bigg]=0,\label{7**}
\end{eqnarray}
Again we transform the above equation in terms of shape function
$\beta(r)$ and employ the numerical approach to show the behavior of
flaring out condition and asymptotic flatness. The left plot of Fig.
\textbf{10} shows $\beta(r)$ as increasing function of $r$. In this
case, throat is located at $r=0.0093117$ with $\beta(r_0)=r_0$ and
$\beta'(r_0)<1$, the behavior of $\beta'(r)$ is shown in right plot
of Fig. \textbf{10}. Moreover, Fig. \textbf{11} shows that our
solutions satisfy the flaring out condition but this solution does
not satisfy the asymptotically flat condition, i.e.,
$\beta(r)/r\rightarrow0$ as $r\rightarrow\infty$. The qualitative
behavior of $\rho$ and $\rho+p_t$ is shown in Fig. \textbf{12}.
Here, we find that the WEC and NEC are not satisfied, so in this
case a realistic wormhole is not possible. Hence, the effective
curvature contributions in the form of exotic matter help to sustain
the wormhole solution.
\begin{figure}\centering
\epsfig{file=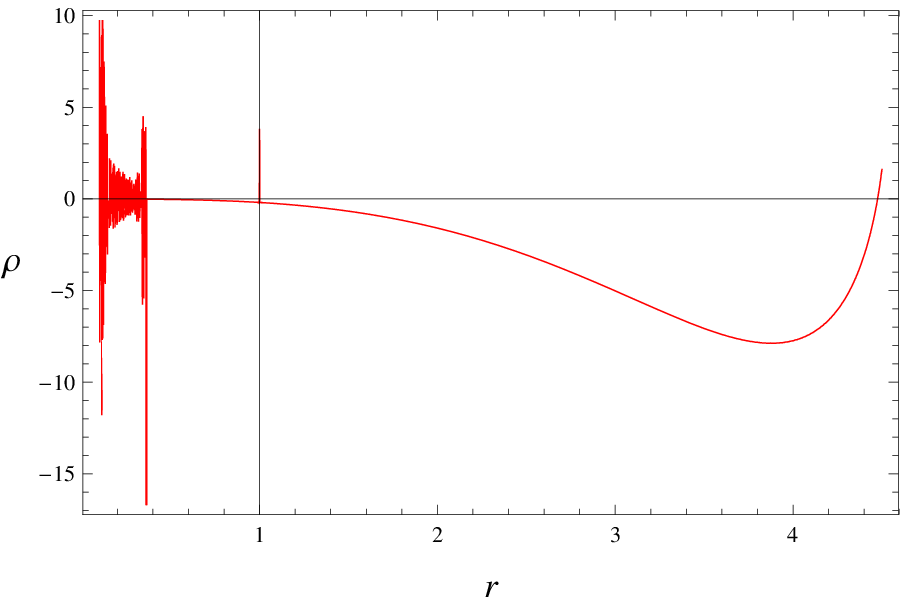, width=0.45\linewidth, height=2in}
\epsfig{file=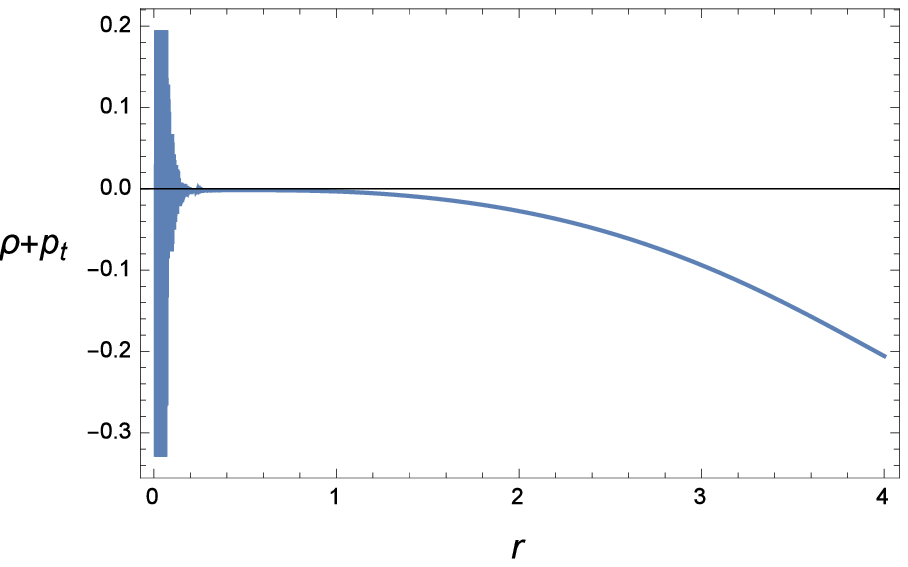, width=0.45\linewidth, height=2in}
\caption{Evolution of $\rho$ and $\rho+p_t$.}
\end{figure}

\section{Summary}

In GR, wormhole solutions contain a fundamental ingredient, that is,
the violation of energy condition in a given space-time. It is taken
into consideration that one may impose the principle of modified
Einstein field equation by effective stress energy momentum tensor
threading the wormholes satisfy the energy conditions and the higher
order curvature derivative terms can support geometries of the
non-standard wormholes. In this manuscript, we have investigated
whether the ordinary matter can support wormholes in $f(R,T)$
modified gravity. For this purpose, we have examined the behavior of
energy conditions, i.e., WEC and NEC, for three different fluids:
barotropic, isotropic and anisotropic fluids in separate cases.

In literature, it is pointed out that the theoretical advances in
the last decades indicate that pressures within highly compact
astrophysical objects are anisotropic, i.e., radial pressure $p_r$
is not equal to tangential pressure $p_t$ in such objects.
Anisotropic matter is more general case than isotropic/barotropic
case, so it is interesting to examine the existence of wormhole
using such matter contents. In this paper, we examined the existence
of wormhole solutions using different matter sources. In literature,
different techniques have been used to discuss wormhole solutions.
One technique is to consider shape function and explore the behavior
of energy conditions and the other technique is to calculate shape
function by taking some assumption for matter ingredients. In this
paper, we are using $f(R,T)$ gravity that involves coupling of
matter and Ricci scalar, therefore the resulting equations are quite
complicated being highly non-linear with six unknowns namely
$\rho,~p_t,~p_r,~b,~a$ and $f(R)$, therefore we should take some
assumptions.

In anisotropic case, it is very difficult to explore the form of
shape function from field equations, therefore we explore the
behavior of energy condition bounds to check possible existence of
wormholes by assuming a viable form of shape function. In other two
cases, that is barotropic and isotropic matter sources, the
equations are less complicated, therefore we have explored the
physical behavior of shape function also. Basically, our purpose is
to check whether the coupling of Ricci scalar with matter field can
support to existence of wormhole geometries in such theory. In order
to discuss wormhole geometries, we have taken some viable
conditions. In all cases, we have assumed a well-defined $f(R)$
model defined as $f(R,T)=f_{1}(R)+{\lambda}T$, where $\lambda\in
\mathbb{R}$. In anisotropic case, we have discussed the existence of
wormhole solutions by taking a particular choice of shape function.
Whereas for the other two cases, we have solved the field equation
numerically to investigate the behavior of shape function. In case
of anisotropic fluid, the behavior of energy constraints have been
discussed for two cases of coupling parameter: $\lambda>0$ and
$\lambda<-1$. For $\lambda<-1$, it is observed that the WEC is valid
for positive values of $\alpha$, while the small validity regions
can be found when $\alpha<0$. For $\lambda<-1$, it is observed that
the validity regions for WEC can be increased by taking large values
of $\alpha$. For $\lambda>0$, the validity of WEC requires negative
range of $\alpha$ whereas small validity regions can be found for
$\alpha>0$.

Firstly, in our obtained result, wormhole solutions exist but these
are not realistic or physically reasonable as one cannot find out
the similarity regions for the compatibility of energy bounds,
although it is mathematically well-defined problem. Our obtained
results are consistent with the works already available in
literature \cite{4n}. Our results are also similar to that obtained
in simple $f(R)$ gravity \cite{17} (that is the case $\lambda=0$).
It is interesting to mention here that in their study, both energy
bounds, i.e., WEC and DEC are violated in anisotropic and isotropic
cases, only in barotropic case, there are some regions where these
conditions are compatible while in our case, only DEC energy bound
is violated for anisotropic case. In other cases, these conditions
remain compatible for some specific ranges of parameters. This
difference of result may be arisen due to the presence of
curvature-matter coupling term.

In literature, the existence of wormhole solutions in
curvature-matter coupled gravity has been discussed by Bertolami and
Ferreira \cite{6n} and Garcia and Lobo \cite{7n}. In both these
studies, they have presented a very restricted analysis in this
sense that they have used linear functions as $f(R)$ model and also
very specific ranges of free parameters have been discussed. They
showed that obtained wormhole solutions are well-behaved satisfying
DEC when $\lambda$ is positive and increasing. It is interesting to
mention here that our results are more comprehensive than these
previous works as we have explored the behavior of involved
functions and existence of wormhole by taking all possible ranges of
the involved parameters (specifically $\lambda$ can take any value).
Furthermore, we have used the Starobinsky model that represents
$R^n$ extension ($n\geq3$) instead of linear functions. Garcia and
Lobo \cite{8n} also discussed the wormhole existence in
curvature-matter coupling gravity by taking linear $f(R)$ model with
positive increasing ansatz for density which is not a physically
reasonable choice for density on cosmological ground (as it should
be a decreasing function)

We have also investigated the equilibrium picture of found wormhole
solutions with anisotropic matter in this gravity. It is seen that
the wormhole solutions are stable as the equilibrium condition
involving hydrostatic and anisotropic forces is satisfied. In the
case of barotropic and isotropic fluids, we have explored the
dynamics of shape function by solving equations numerically. From
the graphical illustrations of shape function, it is seen that in
isotropic case, all the necessary conditions like asymptotically
flatness and flaring out constraint are satisfied which indicates
that the obtained micro wormhole is realistic and viable. While in
case of barotropic fluid, the asymptotic flatness condition is
incompatible therefore, a realistic wormhole solution does not
exist. It is interesting to to find wormhole solutions without
exotic matter by considering some other different $f(R,T)$ models in
this gravity.


\begin{thebibliography}{30}

\bibitem{1} Perlmutter, S. et al.: Astrophys. J. \textbf{483}(1997)565;
Perlmutter, S. et al.: Nature \textbf{391}(1998)51; Perlmutter, S.
et al.: Astrophys. J. \textbf{517}(1999)565.

\bibitem{2} Nojiri, S. and Odintsov, S.D.: Phys. Rept.
\textbf{505}(2011)59.

\bibitem{3} Starobinsky, A.A.: Phys. Lett. B \textbf{91}(1980)99;
Starobinsky, A.A.: JETP Lett. \textbf{86}(2007)157.

\bibitem{4} Ferraro, R. and Fiorini, F.: Phys. Rev. D \textbf{75}(2007)084031;
Zubair, M. Inter. J.  Modern Phys. D \textbf{25}(2016)1650057.

\bibitem{7} Carroll, S. et al.: Phys. Rev. D \textbf{71}(2005)063513;
Cognola.G.: Phys. Rev. D \textbf{73}(2006)084007.

\bibitem{8} Agnese, A.G. and La Camera, M.: Phys. Rev. D
\textbf{51}(1995)2011.

\bibitem{9} Kofinas, G. and Saridakis, N.E.: Phys. Rev. D \textbf{90}(2014)084044.

\bibitem{10} Harko, T. et al.: Phys. Rev. D \textbf{84}(2011)024020.

\bibitem{11} Houndjo, M.J.S. et al.: Int. J. Mod. Phys. D \textbf{21}(2012)1250003.

\bibitem{12} Houndjo, M.J.S. et al.: Int. J. Mod. Phys. \textbf{2}(2012)1250024.

\bibitem{13} Sharif, M. and Zubair, M.: JCAP \textbf{03}(2012)028;
Singh, C.P. and Singh, V.: Gen. Relativ. Gravitt.
\textbf{46}(2014)1696; Mubasher et al.: Eur. Phys. J. C
\textbf{72}(2012)1999; Sharif, M. and Zubair, M.: J. Phys. Soc. Jpn.
\textbf{82}(2013)064001; Shabani, H. and Farhoudi, M.: Phys. Rev. D
\textbf{88}(2013)044048; ibid. Phys. Rev. D \textbf{90}(2014)044031;
Santos, A.F.: Modern Phys. Lett. A \textbf{28}(2013)1350141; M.
Sharif, M. Zubair, Gen. Relativ. Gravit. \textbf{46}(2014)1723;
Zubair, M. and Noureen, I.: Eur. Phys. J. C \textbf{75}(2015)265;
Noureen, I. and Zubair, M.: Eur. Phys. J. C \textbf{75}(2015)62;
Noureen, I. and Zubair, M., Bhatti, A.A. and Abbas, G.: Eur. Phys.
J. C \textbf{75}(2015)323; Alvarenga et al.: Phys. Rev. D
\textbf{87}(2013)103526;  Baffou et al. Astrophys. Space Sci.
\textbf{356}(2015)173; Shamir, M.F.: Eur. Phys. J. C
\textbf{75}(2015)354; Moraes, P.H.R.S.: Eur. Phys. J. C
\textbf{75}(2015)168; Zubair, M. et al.: Astrophys Space Sci.
\textbf{361}(2016)8; Zubair, M. and Syed M. Ali Hassan.: Astrophys
Space Sci. \textbf{361}(2016)149; Shahbani, H.: arXiv:1604.04616v1.

\bibitem{15} Einstein, A. and Rosen, N.: Phys. Rev. \textbf{48}(1935)73.

\bibitem{16} Lobo, F.S.N. and Oliveira, M.A.: Phys. Rev. D \textbf{80}(2009)104012.

\bibitem{17} Sharif, M. and Zahra, Z.: Astrophys. Space Sci. \textbf{348}(2013)275.

\bibitem{18} Morris, M.S. and Thorne, K.S.: Am. J. Phys. \textbf{56}(1988)395; Morris, M.S., Throne, K.S. and Yurtseve, U.: Phys.
Rev. Lett. \textbf{61}(1988)1446.

\bibitem{19} Azizi, T.: Int. J. Theor. Phys. \textbf{52}(2013)3486.

\bibitem{28} Landau, L.D. and Lifshitz, E.M.: \emph{The Classical Theory of Fields} (Butterworth-Heinemann, 2002).

\bibitem{19*}Alvarenga et al.: J. Modern Phys. 4(2013)130; Sharif, M. and Zubair, M.: J. Phys. Soc. Jpn. \textbf{82} (2013)014002.

\bibitem{19**} Ozkan, M. and Pang, Y.: Class. Quantum Grav. \textbf{31}, 205004 (2014).

\bibitem{2s} Goswami, U.D. and Deka, K.: Int. J. Mod. Phys. D \textbf{22}(2013)1350083.

\bibitem{3s} Frolov, A.V.: Phys. Rev. Lett. \textbf{101}(2008)061103.

\bibitem{4s}Hu, W. and Sawicki, I.: Phys. Rev. D \textbf{76}(2007)064004.

\bibitem{20} Jamil, M., Momeni, D. and Myrzakulov, R.: Eur. Phys. J. C \textbf{73}(2013)2267.

\bibitem{21} Boehmer, C.G., Harko, T.and Lobo, F.S.N.: Phys. Rev. D \textbf{85}(2012)044033.

\bibitem{22} Pavlovic, P. and Sossich, M.: Eur. Phys. J. C \textbf{75}(2015)117.

\bibitem{23} Dent, J.B. , Dutta, S. and Saridakis, E.N.: J. Cosmol. Astropart. Phys. \textbf{009}(2011)1101;
Sotiriou, T.P., Li, B. and Barrow, J.D.: Phys. Rev. D
\textbf{83}(2011)104030; Li, B., Sotiriou, T.P. and Barrow, J.D.:
Phys. Rev. D \textbf{83}(2011)104017; Zhang, Y., Li, H., Gong, Y.
and Zhu, Z.H.: J. Cosmol. Astropart. Phys. \textbf{015}(2011)1107;
Bhattacharya, S. and Chakraborty, S.: arXiv:1506.03968v2.

\bibitem{4n} Mehdizadeh, M.R., Zangeneh, M.K. and Lobo, F.S.N.: Phys. Rev. D
\textbf{91}(2015)084004; Jamil, M., Momeni, D. and Myrzakulov, R.:
Eur. Phys. J. C \textbf{73}(2013)2267;

\bibitem{6n} Bertolami, O. and Ferreira, R.Z.: Phys. Rev. D
\textbf{85}(2012)104050.

\bibitem{7n} Garcia, N.M. and Lobo, F.S.N.: Phys. Rev. D
\textbf{82}(2010)104018.

\bibitem{8n} Garcia, N.M. and Lobo, F.S.N.: Class. Quantum Grav. \textbf{28}(2011)085018.

\end{thebibliography}
\end{document}